\documentclass[10pt,reqno]{article}
\usepackage{graphicx}
\usepackage{indentfirst,csquotes}

\usepackage{amssymb,amsthm,amsmath}
\usepackage{xcolor,paralist,hyperref,titlesec,fancyhdr,etoolbox}

\usepackage{graphicx}
\usepackage[numbers]{natbib}

\usepackage{here}

\usepackage{times}
\usepackage{helvet}
\usepackage{courier}
\usepackage{comment}
\usepackage{booktabs}
\usepackage{adjustbox}
\usepackage{threeparttable}

\usepackage{subcaption}

\usepackage[most]{tcolorbox}

\tcbset{
  beamerblock/.style={
    enhanced,
    colback=blue!7,        
    colframe=blue!50!black,
    boxrule=0.6pt,         
    arc=1.2mm,             
    leftrule=0pt, rightrule=0pt, toprule=0pt, bottomrule=0pt,
    left=6pt,right=6pt,top=6pt,bottom=6pt,
    breakable              
  }
}

\usepackage{comment}
\usepackage{booktabs}
\usepackage{threeparttable}

\usepackage{tikz}
\usetikzlibrary{arrows.meta,decorations.pathmorphing,positioning,calc,fit,backgrounds}

\tikzset{
  dep/.style   = {-{Stealth[length=2.1mm]}, thick},
  indep/.style = {-{Stealth[length=2.1mm]},
                  decorate,
                  decoration={snake,amplitude=0.45mm,segment length=1.2mm}},
  lab/.style   = {anchor=east, inner sep=2pt},
  every node/.style = {font=\normalsize},
  highlightblock/.style={
    draw=blue!50!black,
    fill=blue!7,
    line width=0.6pt,
    rounded corners=1.2mm,
    inner sep=6pt
  }
}

\usepackage{amsmath,amssymb,amsfonts,amsthm}
\theoremstyle{plain}
\newtheorem{thm}{Theorem}
\newtheorem*{thm*}{Theorem}

\theoremstyle{plain}
\newtheorem{cor}{Corollary}
\newtheorem*{cor*}{Corollary}

\theoremstyle{plain}

\newtheorem{dfn*}{Definition}

\theoremstyle{plain}

\newtheorem{asm*}{Assumption}

\usepackage{cleveref}

\hypersetup{
    colorlinks=true,
    linkcolor=blue,
    filecolor=blue,
    urlcolor=blue,
    citecolor=blue
}

\usepackage{lipsum}

\begin{document}
 \crefname{figure}{Figure}{Figures}
 \crefname{table}{Table}{Tables}
\title{Convergence Rate of Efficient MCMC with Ancillarity-Sufficiency Interweaving Strategy for Panel Data Models} 
\author{
  Makoto Nakakita\textsuperscript{\rm 1}\thanks{Corresponding author.}, 
  Tomoki Toyabe\textsuperscript{\rm 2},
  Teruo Nakatsuma\textsuperscript{\rm 3},
  Takahiro Hoshino\textsuperscript{\rm 1,3}
  \\
  \textsuperscript{\rm 1}Center for Advanced Intelligence Project, RIKEN \\
  \textsuperscript{\rm 2}Faculty of Economics, Kanazawa Gakuin University \\
  \textsuperscript{\rm 3}Faculty of Economics, Keio University \\
  makoto.nakakita@riken.jp,
}
\date{}
\maketitle

\let\thefootnote\relax

\begin{abstract}
Improving Markov chain Monte Carlo algorithm efficiency is essential for enhancing computational speed and inferential accuracy in Bayesian analysis. These improvements can be effectively achieved using the ancillarity--sufficiency interweaving strategy (ASIS), an effective means of achieving such gains. Herein, we provide the first rigorous theoretical justification for applying ASIS in Bayesian hierarchical panel data models. Asymptotic analysis demonstrated that when the product of prior variance of unobserved heterogeneity and cross-sectional sample size $N$ is sufficiently large, the latent individual effects can be sampled almost independently of their global mean. This near-independence accounts for ASIS's rapid mixing behavior and highlights its suitability for modern ``tall'' panel datasets. We derived simple inequalities to predict which conventional data augmentation scheme---sufficient augmentation (SA) or ancillary augmentation (AA)---yields faster convergence. By interweaving SA and AA, ASIS achieves optimal geometric rate of convergence and renders the Markov chain for the global mean parameter asymptotically independent and identically distributed. Monte Carlo experiment confirm that this theoretical efficiency ordering holds even for small panels (e.g., $N=10$). These findings confirm the empirical success of ASIS application across finance, marketing, and sports, laying the groundwork for its extension to models with more complex covariate structures and non-Gaussian specifications.
\end{abstract}

\bigskip

\section{Introduction}
\label{sec:introduction}
The scope and applicability of Bayesian panel data analysis have expanded with advancements in Markov chain Monte Carlo (MCMC) methods, particularly the Metropolis--Hastings algorithm \cite{metropolis1953equation,hastings1970monte} and Gibbs sampler \cite{geman1984stochastic}; a comprehensive review on Bayesian panel data analysis is provided by \cite{chibPanelDataModeling2008}.

The data augmentation (DA) algorithm \cite{tannerCalculationPosteriorDistributions1987} extends the observed dataset by introducing latent variables to handle incomplete data. Since its introduction, DA has formed the basis for various MCMC extensions and continue to evolve in theoretical and applied contexts.
A single statistical model within a DA framework can have multiple equivalent formulations. Identifying the most computationally efficient formulation---known as the reparametrization problem---is widely studied in Bayesian hierarchical models \cite{hills1992parameterization,gelfand1996efficient,roberts1997updating,papaspiliopoulos2003non,papaspiliopoulos2007general}.
The ancillarity--sufficiency interweaving strategy (ASIS) \cite{yuCenterNotCenter2011} addresses this challenge by alternating between two parameterizations during sampling, rather than relying on a single formulation.
Particularly, ASIS interleaves sufficient augmentation (SA or centered parameterization) and ancillary augmentation (AA or noncentered parameterization). Each scheme exhibits a convergence trade-off: when one converges quickly, the other often slows down \cite{gelfandEfficientParametrisationsNormal1995}. ASIS combines both schemes to achieve superior mixing and convergence than either scheme used alone.

Since its introduction, ASIS has been widely adopted and extended across various domains. Notable early developments include the residual augmentation framework \cite{xu2013thank}, which reinterprets interweaving via residual-based transformations. In financial econometrics, \cite{kastner2014ancillarity} incorporated ASIS into stochastic volatility models, and \cite{kastner2017efficient} later generalized it to multivariate factor stochastic volatility models using shallow and deep interweaving schemes. ASIS has also been used for analyzing financial time-series data in recent studies \cite{Li_2020a,Li_2020b,Nakakita_2021,Toyabe_2024,Nakakita_2025}.

Despite advances in MCMC techniques, the application of Bayesian hierarchical modeling in panel data frameworks remains underdeveloped.
Although ASIS has already been applied in panel data contexts \cite[e.g.]{Saito_2024,Nakakita_2024}, a rigorous theoretical foundation for its use in is lacking.
In addition, strong correlations between random and fixed effects often arise, and treating them as separate blocks can worsen the mixing of the MCMC chain \cite{gelfandEfficientParametrisationsNormal1995}.
Herein, we demonstrate that, under certain conditions, ASIS enables the sampling of unobserved heterogeneity $\alpha$ within panel data frameworks almost independently of the global mean $\mu_\alpha$.

Our primary finding is that when the product of the prior variance of unobserved heterogeneity and the sample size must be sufficiently large, latent heterogeneity effectively decouples from the global parameters. In such cases, the ASIS sampler approximates conditional independence and enables faster mixing than either SA or AA alone. This result highlights ASIS’s compatibility with the growing availability of large-scale (``tall'') panel datasets.
We also derive approximate conditions---within the conventional panel data framework and for sufficiently large sample sizes---under which one can predict whether SA or AA will converge more rapidly.

\section{Convergence Rate of Data Augmentation}
Consider two alternative parameterizations of a panel data regression model: SA and AA. \footnote {As shown in the section of real data analysis, a regression function $\mathbf{x}_{it}^{\intercal}\boldsymbol{\beta}$, where $\mathbf{x}_{it}$ is a vector of covariates and $\boldsymbol{\beta}$ is the corresponding coefficient vector, can be incorporated into equations \eqref{sa} or \eqref{aa} by redefining $y_{it}$ as $y_{it}-\mathbf{x}_{it}^{\intercal}\boldsymbol{\beta}$.}.
\begin{align}
  \label{sa}
  \text{(SA)}\quad
  y_{it} &= \alpha_{i} + \epsilon_{it},\quad \alpha_{i} \sim \mathcal{N}(\mu_{\alpha},\sigma_{\alpha}^{2}), \\
  \label{aa}
  \text{(AA)}\quad
  y_{it} &= \mu_{\alpha} + \tilde{\alpha}_{i} + \epsilon_{it},\quad \tilde{\alpha}_{i} \sim \mathcal{N}(0,\sigma_{\alpha}^{2}), \\
  \epsilon_{it} &\sim \mathcal{N}(0,\sigma_{\epsilon}^{2}),\quad
  i =1,\dots, N,\quad t = 1,\dots, T, \nonumber
\end{align}
where $y_{it}$ denotes the observation for Individual $i$ at a time $t$, $\alpha_{i}$ represents the individual-specific mean of $y_{it}$, and $\tilde{\alpha}_{i}$ is the deviation of $\alpha_{i}$ from the global mean (hyperparameter) $\mu_{\alpha}$.

\begin{figure}[htbp]
\centering
\begin{tikzpicture}[node distance=1.2cm and 0.9cm, >=Stealth, scale=1, every node/.style={scale=0.85}]

\node at (-5.5,2.2) {sufficient augmentation (SA)};
\node at (-5.5,1.8) {$(\mu_\alpha, \sigma_\alpha^2)$};

\node (alpha1) at (-7.3,0.5) {$\alpha_1$};
\node (alpha2) at (-6.5,0.5) {$\alpha_2$};
\node (alpha3) at (-5.7,0.5) {$\alpha_3$};
\node (dots1) at (-4.9,0.5) {$\dots$};
\node (alphaT) at (-4.1,0.5) {$\alpha_T$};

\node (y1) at (-7.3,-1.0) {$y_1$};
\node (y2) at (-6.5,-1.0) {$y_2$};
\node (y3) at (-5.7,-1.0) {$y_3$};
\node (dots2) at (-4.9,-1.0) {$\dots$};
\node (yT) at (-4.1,-1.0) {$y_T$};

\foreach \i/\j in {alpha1/alpha2, alpha2/alpha3, alpha3/dots1, dots1/alphaT} {
    \draw[->] (\i) -- (\j);
}
\foreach \i/\j in {alpha1/y1, alpha2/y2, alpha3/y3, alphaT/yT} {
    \draw[->] (\i) -- (\j);
}
\foreach \i in {alpha1,alpha2,alpha3,alphaT} {
    \draw[->] (-5.5,1.6) -- (\i);
}

\node at (-1.2,2.2) {\small ancillary augmentation (AA)};
\node at (-1.2,1.8) {$(\sigma_\alpha^2)$};

\node (balpha1) at (-2.5,0.5) {$\tilde{\alpha}_1$};
\node (balpha2) at (-1.7,0.5) {$\tilde{\alpha}_2$};
\node (balpha3) at (-0.9,0.5) {$\tilde{\alpha}_3$};
\node (dots3) at (-0.1,0.5) {$\dots$};
\node (balphaT) at (0.7,0.5) {$\tilde{\alpha}_T$};

\node (y1b) at (-2.5,-1.0) {$y_1$};
\node (y2b) at (-1.7,-1.0) {$y_2$};
\node (y3b) at (-0.9,-1.0) {$y_3$};
\node (dots4) at (-0.1,-1.0) {$\dots$};
\node (yTb) at (0.7,-1.0) {$y_T$};

\node (mu) at (-1.2,-2.0) {$\mu_\alpha$};

\foreach \i/\j in {balpha1/balpha2, balpha2/balpha3, balpha3/dots3, dots3/balphaT} {
    \draw[->] (\i) -- (\j);
}
\foreach \i/\j in {balpha1/y1b, balpha2/y2b, balpha3/y3b, balphaT/yTb} {
    \draw[->] (\i) -- (\j);
}
\foreach \i in {balpha1,balpha2,balpha3,balphaT} {
    \draw[->] (-1.2,1.6) -- (\i);
}
\foreach \i in {y1b,y2b,y3b,yTb} {
    \draw[->] (mu) -- (\i);
}
\end{tikzpicture}
\caption{Comparison of parameter dependence between sufficient augmentation (SA) and ancillary augmentation (AA).}
\end{figure}

 Suppose the prior distribution of $\mu_{\alpha}$ is $\mathcal{N}(\varphi_{\alpha},\tau_{\alpha}^{2})$, and $\sigma_{\epsilon}^{2}$ and $\sigma_{\alpha}^{2}$ are known.
 
Let $\mathcal{D}=(y_{11},\dots,y_{NT})$ denote the observed data, $\alpha=(\alpha_{1},\dots,\alpha_{N})$ denote the vector of individual-specific effects, and $\tilde{\alpha}=(\tilde{\alpha}_{1},\dots,\tilde{\alpha}_{N})$ denote the vector of deviations from the global mean.
The Gibbs sampling algorithm under each parameterization is then derived as follows:
\begin{align}
  &\text{(SA)}\nonumber\\
  \label{sa_gibbs_alpha}
  &\alpha_{i}|\mu_{\alpha},\mathcal{D} \sim \mathcal{N}\left(\frac{\sigma_{\epsilon}^{-2}\sum_{t=1}^{T}y_{it} + \sigma_{\alpha}^{-2}\mu_{\alpha}} {\lambda_{\alpha}}, \frac{1}{\lambda_{\alpha}}\right), \\
  \label{sa_gibbs_mu}
  &\mu_{\alpha}|\alpha,\mathcal{D} \sim \mathcal{N}\left(\frac{\sigma_{\alpha}^{-2}\sum_{i=1}^{N}\alpha_{i} + \tau_{\alpha}^{-2}\varphi_{\alpha}}{\lambda_{\mu SA}}, \frac{1}{\lambda_{\mu SA}}\right), \\
  &\text{(AA)}\nonumber\\
  \label{aa_gibbs_alpha}
  &\tilde{\alpha}_{i}|\mu_{\alpha},\mathcal{D} \sim \mathcal{N}\left(\frac{\sigma_{\epsilon}^{-2}\sum_{t=1}^{T}(y_{it}-\mu_{\alpha})}{\lambda_{\alpha}}, \frac{1}{\lambda_{\alpha}}\right), \\
  \label{aa_gibbs_mu}
  &\mu_{\alpha}|\tilde{\alpha},\mathcal{D} \sim \mathcal{N}\left(\frac{\sigma_{\epsilon}^{-2}\sum_{i=1}^{N}\sum_{t=1}^{T}(y_{it}-\tilde{\alpha}_{i}) + \tau_{\alpha}^{-2}\varphi_{\alpha}}{\lambda_{\mu AA}}, \frac{1}{\lambda_{\mu AA}}\right),
\end{align}
\begin{align*}
\lambda_{\alpha} = \sigma_{\epsilon}^{-2}T + \sigma_{\alpha}^{-2}, 
\quad \lambda_{\mu SA} = \sigma_{\alpha}^{-2}N + \tau_{\alpha}^{-2},
\quad \lambda_{\mu AA} = \sigma_{\epsilon}^{-2}NT + \tau_{\alpha}^{-2}.
\end{align*}

Introduce $z_{0},z_{1},\dots,z_{N}\sim \mathcal{N}(0,1)$.
Then, \eqref{sa_gibbs_alpha} and \eqref{sa_gibbs_mu} are equivalent to
\begin{align}
  \label{sa_gibbs_alpha_alt}
  \alpha_{i} &= \frac{\sigma_{\epsilon}^{-2}\sum_{t=1}^{T}y_{it} + \sigma_{\alpha}^{-2}\mu_{\alpha}}{\lambda_{\alpha}} + \frac{z_{i}}{\sqrt{\lambda_{\alpha}}}, \quad i=1,\dots, N, \\
  \label{sa_gibbs_mu_alt}
  \mu_{\alpha} &= \frac{\sigma_{\alpha}^{-2}\sum_{i=1}^{N}\alpha_{i} + \tau_{\alpha}^{-2}\varphi_{\alpha}}{\lambda_{\mu SA}} + \frac{z_{0}}{\sqrt{\lambda_{\mu SA}}},
\end{align}
respectively. Similarly, equations \eqref{aa_gibbs_alpha} and \eqref{aa_gibbs_mu} can be rearranged as follows:
\begin{align}
  \label{aa_gibbs_alpha_alt}
  \tilde{\alpha}_{i} &= \frac{\sigma_{\epsilon}^{-2}\sum_{t=1}^{T}(y_{it}-\mu_{\alpha})}{\lambda_{\alpha}} + \frac{z_{i}}{\sqrt{\lambda_{\alpha}}},\quad i=1,\dots,N, \\
  \label{aa_gibbs_mu_alt}
  \mu_{\alpha} &= \frac{\sigma_{\epsilon}^{-2}\sum_{i=1}^{N}\sum_{t=1}^{T}(y_{it}-\tilde{\alpha}_{i}) + \tau_{\alpha}^{-2}\varphi_{\alpha}}{\lambda_{\mu AA}} + \frac{z_{0}}{\sqrt{\lambda_{\mu AA}}}.
\end{align}
\indent The Markov process for $\mu_{\alpha}$ in the Gibbs sampler under the SA parameterization is then derived from equations \eqref{sa_gibbs_alpha_alt} and \eqref{sa_gibbs_mu_alt} as follows:
\begin{align}
  \mu_{\alpha}^{(r+1)} &= \frac{\sigma_{\alpha}^{-2}}{\lambda_{\mu SA}} \sum_{i=1}^{N}\left(\frac{\sigma_{\epsilon}^{-2}\sum_{t=1}^{T}y_{it}
  + \sigma_{\alpha}^{-2}\mu_{\alpha}^{(r)}}{\lambda_{\alpha}} + \frac{z_{i}}{\sqrt{\lambda_{\alpha}}}\right) + \frac{\tau_{\alpha}^{-2}\varphi_{\alpha}}{\lambda_{\mu SA}} + \frac{z_{0}}{\sqrt{\lambda_{\mu SA}}} \nonumber \\
  &= \frac{\sigma_{\alpha}^{-2}}{\lambda_{\mu SA}}\left(\frac{\sigma_{\alpha}^{-2}N}{\lambda_{\alpha}}\mu_{\alpha}^{(r)} + \frac{\sigma_{\epsilon}^{-2}\sum_{i=1}^{N}\sum_{t=1}^{T}y_{it}}{\lambda_{\alpha}} \right. \left. + \frac{\sum_{i=1}^{N}z_{i}}{\sqrt{\lambda_{\alpha}}}\right) + \frac{\tau_{\alpha}^{-2}\varphi_{\alpha}}{\lambda_{\mu SA}} + \frac{z_{0}}{\sqrt{\lambda_{\mu SA}}} \nonumber \\
  &= \frac{\sigma_{\alpha}^{-2}}{\sigma_{\alpha}^{-2} + \tau_{\alpha}^{-2}/N}\left(\frac{\sigma_{\alpha}^{-2}}{\lambda_{\alpha}}\mu_{\alpha}^{(r)} + \frac{\sigma_{\epsilon}^{-2}T\bar{y}}{\lambda_{\alpha}} + \frac{\bar{z}}{\sqrt{\lambda_{\alpha}}}\right) + \frac{\tau_{\alpha}^{-2}\varphi_{\alpha}}{\lambda_{\mu SA}} + \frac{z_{0}}{\sqrt{\lambda_{\mu SA}}},\quad r = 1,\dots,R, \label{sa_gibbs_mu_markov}
\end{align}
where $\bar{y} = \frac{1}{NT}\sum_{i=1}^{N}\sum_{t=1}^{T}y_{it}$ and $\bar{z} = \frac{1}{N}\sum_{i=1}^{N}z_{i}$.
Similarly, the Markov process for $\mu_{\alpha}$ in the Gibbs sampler under the AA parameterization is derived from equations \eqref{aa_gibbs_alpha_alt} and \eqref{aa_gibbs_mu_alt} as follows:
\begin{align}
  \label{aa_gibbs_mu_markov}
  \mu_{\alpha}^{(r + 1)} &=
  \frac{\sigma_{\epsilon}^{-2}}{\lambda_{\mu AA}} \sum_{i=1}^{N}\sum_{t=1}^{T} \left(y_{it}-\frac{\sigma_{\epsilon}^{-2}\sum_{t=1}^{T}(y_{it}-\mu_{\alpha}^{(r)})}
  {\lambda_{\alpha}} - \frac{z_{i}}{\sqrt{\lambda_{\alpha}}}\right) +\frac{\tau_{\alpha}^{-2}\varphi_{\alpha}}{\lambda_{\mu AA}} + \frac{z_{0}}{\sqrt{\lambda_{\mu AA}}} \nonumber \\
  &=
  \frac{\sigma_{\epsilon}^{-2}}{\lambda_{\mu AA}}\left(\frac{\sigma_{\epsilon}^{-2}NT^{2}}{\lambda_{\alpha}}\mu_{\alpha}^{(r)} + NT\bar{y} - \frac{\sigma_{\epsilon}^{-2}NT^{2}\bar{y}}{\lambda_{\alpha}} - \frac{NT\bar{z}}{\sqrt{\lambda_{\alpha}}}
  \right) +\frac{\tau_{\alpha}^{-2}\varphi_{\alpha}}{\lambda_{\mu AA}} + \frac{z_{0}}{\sqrt{\lambda_{\mu AA}}} \nonumber \\
  &= \frac{\sigma_{\epsilon}^{-2}}{\sigma_{\epsilon}^{-2} + \tau_{\alpha}^{-2}/NT}\left(
  \frac{\sigma_{\epsilon}^{-2}T}{\lambda_{\alpha}}\mu_{\alpha}^{(r)} + \frac{\sigma_{\alpha}^{-2}\bar{y}}{\lambda_{\alpha}} - \frac{\bar{z}}{\sqrt{\lambda_{\alpha}}}\right) +\frac{\tau_{\alpha}^{-2}\varphi_{\alpha}}{\lambda_{\mu AA}} + \frac{z_{0}}{\sqrt{\lambda_{\mu AA}}},\quad r=1,\dots,R.
\end{align}

When $\tau_{\alpha}^{2}N$ is sufficiently large,
\begin{align}
  \frac{\sigma_{\alpha}^{-2}}{\sigma_{\alpha}^{-2} + \tau_{\alpha}^{-2}/N} &\rightarrow 1,\quad
  \frac{\tau_{\alpha}^{-2}\varphi_{\alpha}}{\sigma_{\alpha}^{-2}N + \tau_{\alpha}^{-2}} \rightarrow 0, \quad (SA)\nonumber\\ 
   \frac{\sigma_{\epsilon}^{-2}}{\sigma_{\epsilon}^{-2} + \tau_{\alpha}^{-2}/NT} &\rightarrow 1, \quad \frac{\tau_{\alpha}^{-2}\varphi_{\alpha}}{\sigma_{\epsilon}^{-2}NT + \tau_{\alpha}^{-2}} \rightarrow 0 \quad (AA). \nonumber
\end{align}
As $\tau_{\alpha}^{2}N \rightarrow \infty$,  \eqref{sa_gibbs_mu_markov} and \eqref{aa_gibbs_mu_markov} reduce to AR(1) processes:
\begin{align}
  \label{sa_gibbs_mu_markov_approx}
  \mu_{\alpha}^{(r+1)} &\rightarrow \frac{\sigma_{\alpha}^{-2}}{\lambda_{\alpha}}\mu_{\alpha}^{(r)}+\frac{\sigma_{\epsilon}^{-2}T\bar{y}}{\lambda_{\alpha}} + \frac{\bar{z}}{\sqrt{\lambda_{\alpha}}}+ \frac{z_{0}}{\sqrt{\lambda_{\mu SA}}},\\
  \label{aa_gibbs_mu_markov_approx}
  \mu_{\alpha}^{(r+1)} &\rightarrow \frac{\sigma_{\epsilon}^{-2}T}{\lambda_{\alpha}}\mu_{\alpha}^{(r)} + \frac{\sigma_{\alpha}^{-2}\bar{y}}{\lambda_{\alpha}} - \frac{\bar{z}}{\sqrt{\lambda_{\alpha}}} + \frac{z_{0}}{\sqrt{\lambda_{\mu AA}}}.
\end{align}

\begin{thm}
\label{thm:theorem1}
In the approximated Markov processes \eqref{sa_gibbs_mu_markov_approx} and \eqref{aa_gibbs_mu_markov_approx}, the sum of two AR(1) coefficients is constant, i.e.,
\begin{align}
\label{eq:tradeoff}
  \frac{\sigma_{\epsilon}^{-2}T}{\sigma_{\epsilon}^{-2}T + \sigma_{\alpha}^{-2}} 
  + \frac{\sigma_{\alpha}^{-2}}{\sigma_{\epsilon}^{-2}T + \sigma_{\alpha}^{-2}} &= 1.
\end{align}
\end{thm}
This theorem highlights a fundamental trade-off in convergence behavior: if one algorithm converges rapidly (i.e., exhibits good mixing), the other necessarily converges more slowly (i.e., shows stronger persistence).
\begin{cor}
\label{cor:corollary1}
From Theorem \ref{thm:theorem1}, the following trade-off relationship can be easily derived:
\begin{equation}
\label{eq:cor1}
  \begin{cases}
    \text{SA converges faster if}\ \sigma_{\epsilon}^{2} < \sigma_{\alpha}^{2}T; \\
    \text{AA converges faster if}\  \sigma_{\epsilon}^{2} > \sigma_{\alpha}^{2}T.
  \end{cases}
\end{equation}
\end{cor}
This corollary can be directly derived from equation \eqref{eq:tradeoff}. This indicates that the relative convergence rates of SA and AA depend on the length of the time dimension $T$ in the panel data.
Although this result is asymptotic, we subsequently demonstrate via simulations that the ordering established in Corollary 1 persists even for small panel datasets, with as few as $N=10$ individuals.

\section{Why ASIS Improves the Convergence Rate}
The SA$\to$AA ASIS algorithm is summarized below:
\begin{align}
    \alpha_{i}^{(r+0.5)} &= \frac{\sigma_{\epsilon}^{-2}\sum_{t=1}^{T}y_{it} + \sigma_{\alpha}^{-2}\mu_{\alpha}^{(r)}}{\lambda_{\alpha}} + \frac{z_{i}^{(r+0.5)}}{\sqrt{\lambda_{\alpha}}}, \nonumber \\
    \mu_{\alpha}^{(r+0.5)} &= \frac{\sigma_{\alpha}^{-2}\sum_{i=1}^{N}\alpha_{i}^{(r+0.5)} + \tau_{\alpha}^{-2}\varphi_{\alpha}}{\lambda_{\mu SA}} + \frac{z_{0}^{(r+0.5)}}{\sqrt{\lambda_{\mu SA}}}, \nonumber \\
    \tilde{\alpha}_{i}^{(r+0.5)} &= \alpha_{i}^{(r+0.5)} - \mu_{\alpha}^{{(r+0.5)}}, \nonumber \\
    \mu_{\alpha}^{(r+1)} &= \frac{\sigma_{\epsilon}^{-2}\sum_{i=1}^{N}\sum_{t=1}^{T}(y_{it}-\tilde{\alpha}_{i}^{(r+0.5)}) + \tau_{\alpha}^{-2}\varphi_{\alpha}}{\lambda_{\mu AA}} + \frac{z_{0}^{(r+1)}}{\sqrt{\lambda_{\mu AA}}}. \label{asis_sa_aa}
\end{align}
If the approximation in equation \eqref{sa_gibbs_mu_markov_approx} holds, then we obtain:
\begin{align}
  \tilde{\alpha}_{i}^{(r+0.5)} &= \alpha_{i}^{(r+0.5)} - \mu_{\alpha}^{{(r+0.5)}} \nonumber \\
  &= \frac{\sigma_{\epsilon}^{-2}\sum_{t=1}^{T}y_{it} + \sigma_{\alpha}^{-2}\mu_{\alpha}^{(r)}}{\lambda_{\alpha}} + \frac{z_{i}^{(r+0.5)}}{\sqrt{\lambda_{\alpha}}} - \frac{\sigma_{\alpha}^{-2}\mu_{\alpha}^{(r)}}{\lambda_{\alpha}} - \frac{\sigma_{\epsilon}^{-2}T\bar{y}}{\lambda_{\alpha}} - \frac{\bar{z}^{(r+0.5)}}{\sqrt{\lambda_{\alpha}}}- \frac{z_{0}^{(r+0.5)}}{\sqrt{\lambda_{\mu SA}}} \nonumber \\
  &= \frac{\sigma_{\epsilon}^{-2}\sum_{t=1}^{T}(y_{it} - \bar{y})}{\lambda_{\alpha}} + \frac{z_{i}^{(r+0.5)} - \bar{z}^{(r+0.5)}}{\sqrt{\lambda_{\alpha}}} - \frac{z_{0}^{(r+0.5)}}{\sqrt{\lambda_{\mu SA}}}, \label{asis_sa_aa_alpha}
\end{align}
which implies that $\tilde{\alpha}_{i}^{(r+0.5)}$ does not depend on $\mu_{\alpha}^{(r)}$. Thus, $\mu_{\alpha}^{(r+1)}$ in \eqref{asis_sa_aa} is independent of $\mu_{\alpha}^{(r)}$. 

The AA$\to$SA ASIS algorithm is summarized as follows:
\begin{equation}
  \label{asis_aa_sa}
  \begin{split}
  \tilde{\alpha}_{i}^{(r+0.5)} &= \frac{\sigma_{\epsilon}^{-2}\sum_{t=1}^{T}(y_{it}-\mu_{\alpha}^{(r)})}{\lambda_{\alpha}} + \frac{z_{i}^{(r+0.5)}}{\sqrt{\lambda_{\alpha}}}, \\
  \mu_{\alpha}^{(r+0.5)} &= \frac{\sigma_{\epsilon}^{-2}\sum_{i=1}^{N}\sum_{t=1}^{T}(y_{it}-\tilde{\alpha}_{i}^{(r+0.5)}) + \tau_{\alpha}^{-2}\varphi_{\alpha}}{\lambda_{\mu AA}} + \frac{z_{0}^{(r+0.5)}}{\sqrt{\lambda_{\mu AA}}}, \\
  \alpha_{i}^{(r+0.5)} &= \tilde{\alpha}_{i}^{(r+0.5)} + \mu_{\alpha}^{(r+0.5)}, \\
  \mu_{\alpha}^{(r+1)} &= \frac{\sigma_{\alpha}^{-2}\sum_{i=1}^{N}\alpha_{i}^{(r+0.5)} + \tau_{\alpha}^{-2}\varphi_{\alpha}}{\lambda_{\mu SA}} +\frac{z_{0}^{(r+1)}}{\sqrt{\lambda_{\mu SA}}}.
  \end{split}
\end{equation}
If the approximation in equation \eqref{aa_gibbs_mu_markov_approx} holds, then we have:
\begin{align}
  \label{asis_aa_sa_alpha}
  \alpha_{i}^{(r+0.5)} &= \tilde{\alpha}_{i}^{(r+0.5)} + \mu_{\alpha}^{(r+0.5)} \nonumber \\
  &= \frac{\sigma_{\epsilon}^{-2}\sum_{t=1}^{T}(y_{it}-\mu_{\alpha}^{(r)})}{\lambda_{\alpha}} + \frac{z_{i}^{(r+0.5)}}{\sqrt{\lambda_{\alpha}}} + \frac{\sigma_{\epsilon}^{-2}T\mu_{\alpha}^{(r)}}{\lambda_{\alpha}} + \frac{\sigma_{\alpha}^{-2}\bar{y}}{\lambda_{\alpha}} - \frac{\bar{z}^{(r+0.5)}}{\sqrt{\lambda_{\alpha}}} + \frac{z_{0}^{(r+0.5)}}{\sqrt{\lambda_{\mu AA}}} \nonumber \\
  &= \bar{y} + \frac{z_{i}^{(r+0.5)} - \bar{z}^{(r+0.5)}}{\sqrt{\lambda_{\alpha}}} + \frac{z_{0}^{(r+0.5)}}{\sqrt{\lambda_{\mu AA}}}.
\end{align}
As with equation \eqref{asis_sa_aa}, $\mu_{\alpha}^{(r + 1)}$ in equation \eqref{asis_aa_sa} is independent of $\mu_{\alpha}^{(r)}$. 

\begin{thm}
\label{thm:theorem2}
$\{\mu_{\alpha}^{(r)}\}_{r=1}^{R}$ generated by SA$\to$AA and AA$\to$SA ASIS algorithms is approximately an independent and identically distributed (IID) sequence.
\end{thm}
This theorem follows directly from equations \eqref{asis_sa_aa} and \eqref{asis_aa_sa}.
Furthermore, if $\mu_{\alpha}^{(r)}$ is approximately IID, then so is $\alpha_{i}^{(r + 0.5)}$. 
In conclusion, SA$\to$AA and AA$\to$SA ASIS algorithms can generate an approximately IID sequence $\{(\alpha_{1}^{(r)},\dots,\alpha_{N}^{(r)},\mu_{\alpha}^{(r)})\}_{r=1}^{R}$ when the product $\tau_{\alpha}^{2}N$ is sufficiently large.
\cref{fig:sampling_flowbe} shows that $\mu_{\alpha}^{(r + 1)}$ becomes independent of $\mu_{\alpha}^{(r)}$ under the ASIS framework. Therein, the solid arrows denote dependency relationships between $\mu_\alpha$ and $\alpha$, whereas the wavy arrows indicate asymptotic independence.
The blue boxed segment in Figure \ref{fig:sampling_flowbe} highlights a key insight: by generating $\mu_{\alpha}^{(r + 1)}$ via SA and AA steps, $\mu_{\alpha}^{(r + 1)}$ becomes independent of $\mu_{\alpha}^{(r)}$.

\begin{figure*}[tbp]
\begin{tikzpicture}[%
  >=Stealth,
  node distance = 18mm,
  decoration = {snake, amplitude = .6mm, segment length = 2mm},
  every node/.style = {font = \normalsize}
]

\draw (4,0) -- (4,-4);

\node[anchor = west] at (0.4,-2.5) {$\mathrm{SA}\rightarrow\mathrm{AA}\;(\mathrm{ASIS})$};
\node[anchor = west] at (0.4,-3.5) {$\mathrm{AA}\rightarrow\mathrm{SA}\;(\mathrm{ASIS})$};
\node[anchor = west] at (1.4,-0.5) {$\mathrm{SA}$};
\node[anchor = west] at (1.4,-1.5) {$\mathrm{AA}$};

\node (mu0)  at (5,-3)   {$\mu_{\alpha}^{(r)}$};

\node (a1)   at (8,-2.5) {$\alpha^{(r+0.5)}$};
\node (ta1)  at (8,-3.5) {$\tilde{\alpha}^{(r+0.5)}$};

\node (mu05) at (11,-3)   {$\mu_{\alpha}^{(r+0.5)}$};

\node (ta1r)  at (14,-2.5) {$\tilde{\alpha}^{(r+0.5)}$};
\node (a1r) at (14,-3.5) {$\alpha^{(r+0.5)}$};

\node (mu1)  at (17,-3)  {$\mu_{\alpha}^{(r+1)}$};

\draw[->] (mu0) -- (a1);
\draw[->] (mu0) -- (ta1);

\draw[->] (a1)  -- (mu05);
\draw[->] (ta1) -- (mu05);

\draw[->,decorate] (mu05) -- (a1r);
\draw[->,decorate] (mu05) -- (ta1r);

\draw[->] (a1r) -- (mu1);
\draw[->] (ta1r) -- (mu1);

\draw[->,decorate] (a1)  -- (ta1r);
\draw[->,decorate] (ta1) -- (a1r);

\node (mu0c) at (5,-0.5) {$\mu_{\alpha}^{(r)}$};
\node (ahat) at (11,-0.5) {$\alpha^{(r+1)}$};
\node (mu1c) at (17,-0.5) {$\mu_{\alpha}^{(r+1)}$};

\draw[->] (mu0c) -- (ahat);
\draw[->] (ahat) -- (mu1c);

\node (mu0d) at (5,-1.5) {$\mu_{\alpha}^{(r)}$};
\node (t1)   at (11,-1.5) {$\tilde{\alpha}^{(r+1)}$};
\node (mu1d) at (17,-1.5) {$\mu_{\alpha}^{(r+1)}$};

\draw[->] (mu0d) -- (t1);
\draw[->] (t1)   -- (mu1d);

\begin{pgfonlayer}{background}
\node[highlightblock,fit=(a1)(a1r)] {};
\end{pgfonlayer}

\end{tikzpicture}
\caption{Sampling flow for SA, AA, and ancillarity--sufficiency interweaving strategy (ASIS). In SA and AA, the updated global mean $\mu_{\alpha}^{(r + 1)}$ depends on its previous value $\mu_{\alpha}^{(r)}$. Contrarily, in ASIS, $\mu_{\alpha}^{(r + 1)}$ is generated independently of $\mu_{\alpha}^{(r)}$.}
\label{fig:sampling_flowbe}
\end{figure*}

\section{Simulation Study}
The theoretical results were verified via simulations using synthetic data generated under several scenarios described below.
MCMC estimations were conducted across 18 settings for six panel configurations: $(N, T) = (10,10)$, $(10,100)$, $(10,500)$, $(500,10)$, $(500,100)$ and $(500,500)$. 
These included three patterns: Pattern 1, where SA is expected to converge faster; Pattern 2, where AA is expected to converge faster; and Pattern 3, where SA and AA are expected to exhibit comparable rates. 
These expectations were based on the relative convergence derived in Corollary 1. 
Each simulation was run for 10,000 iterations, with a 1,000-iteration burn-in period. 
The results were averaged over 100 independent MCMC runs. 
Because the convergence conditions in Corollary 1 depend on the time dimension $T$, the prior distribution parameters used for each configuration are provided in the footnotes of the corresponding tables.

\begin{figure}[H]
    \centering

    \begin{subfigure}[t]{0.49\textwidth}
        \centering
        \includegraphics[width=\linewidth]{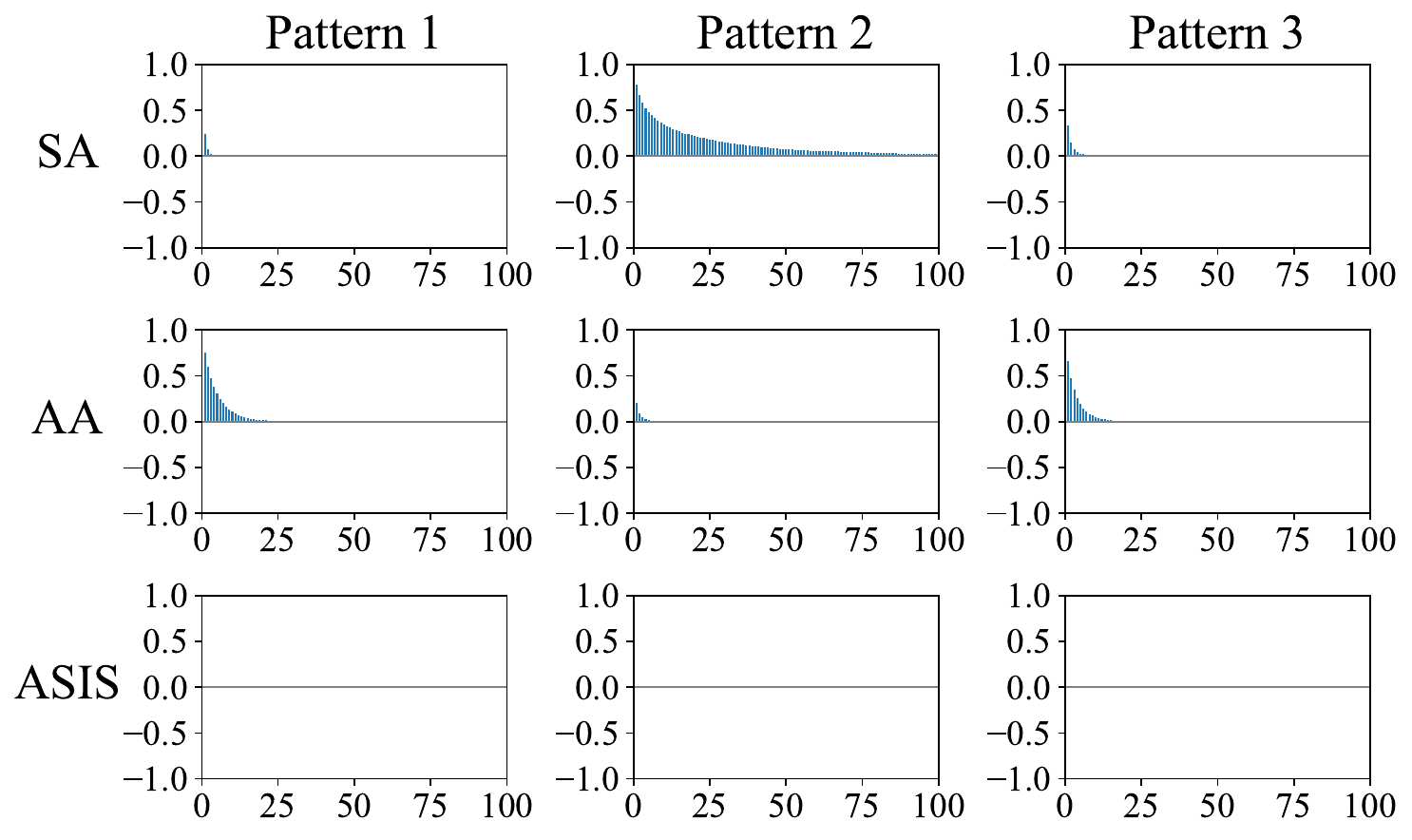}
        \caption{$(N,T) = (10,10)$}
        \label{fig:acf_n10_t10}
    \end{subfigure}
    \hfill
    \begin{subfigure}[t]{0.49\textwidth}
        \centering
        \includegraphics[width=\linewidth]{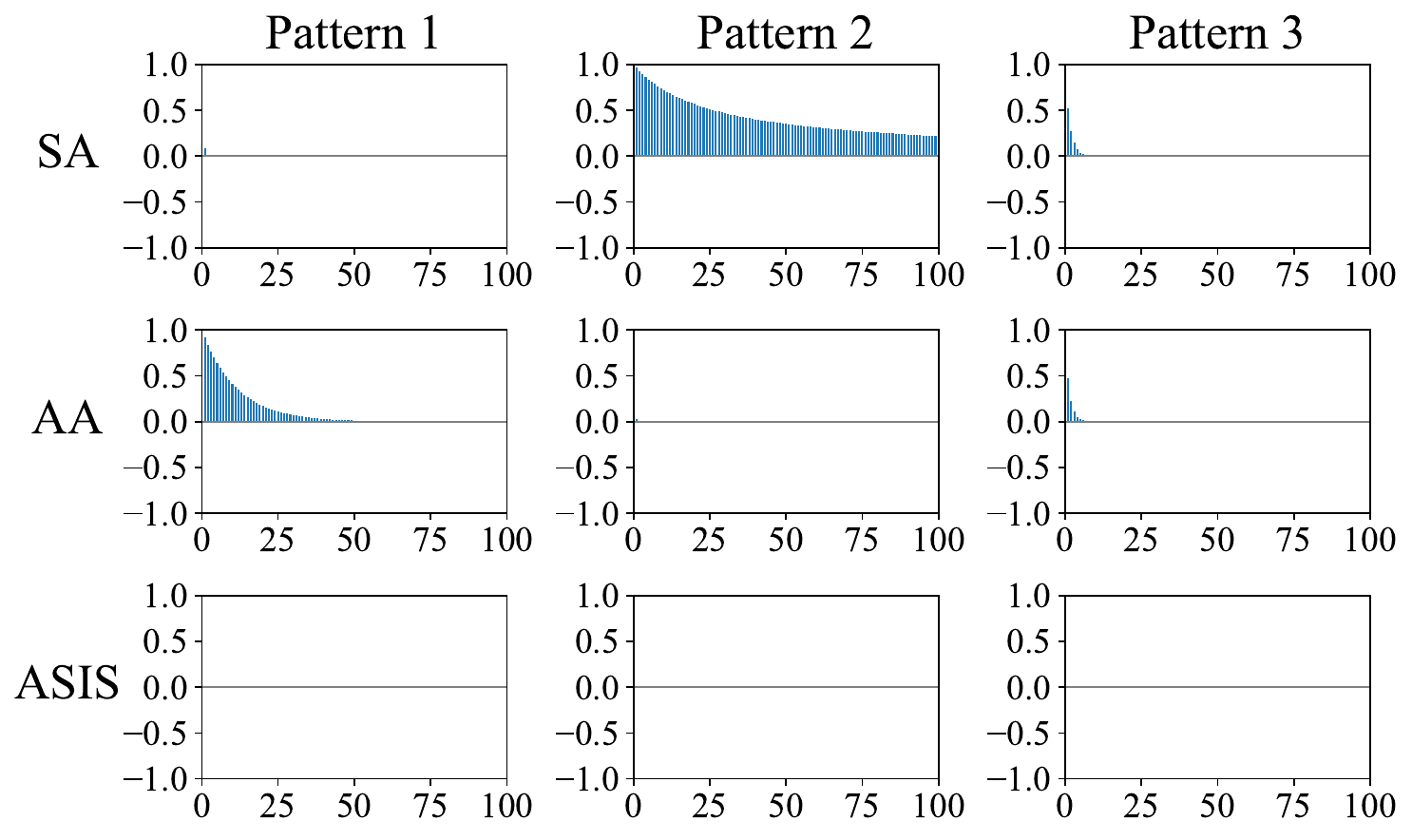}
        \caption{$(N,T) = (500,10)$}
        \label{fig:acf_n500_t10}
    \end{subfigure}

    \begin{subfigure}[t]{0.49\textwidth}
        \centering
        \includegraphics[width=\linewidth]{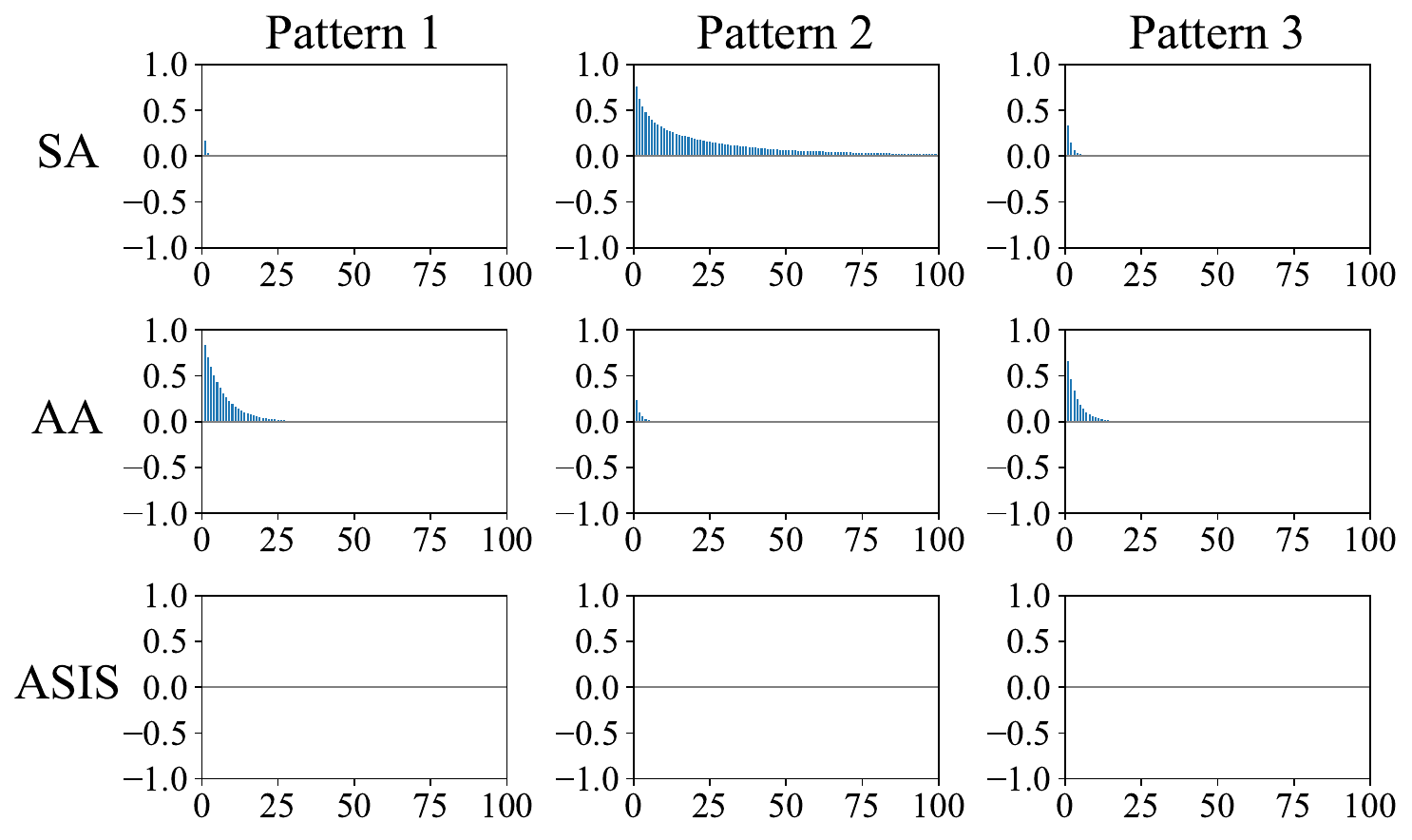}
        \caption{$(N,T) = (10,100)$}
        \label{fig:acf_n10_t100}
    \end{subfigure}
    \hfill
    \begin{subfigure}[t]{0.49\textwidth}
        \centering
        \includegraphics[width=\linewidth]{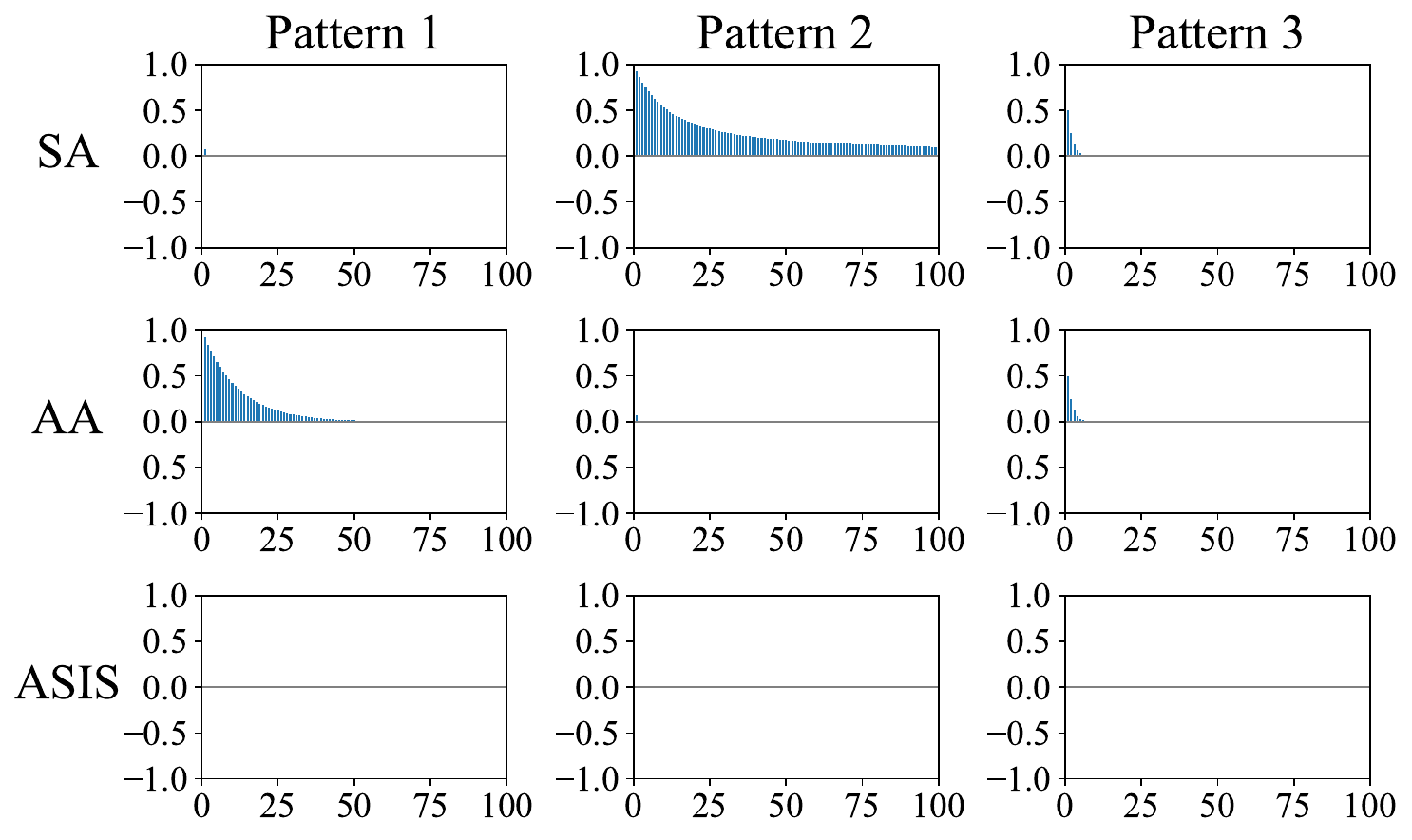}
        \caption{$(N,T) = (500,100)$}
        \label{fig:acf_n500_t100}
    \end{subfigure}

    \begin{subfigure}[t]{0.49\textwidth}
        \centering
        \includegraphics[width=\linewidth]{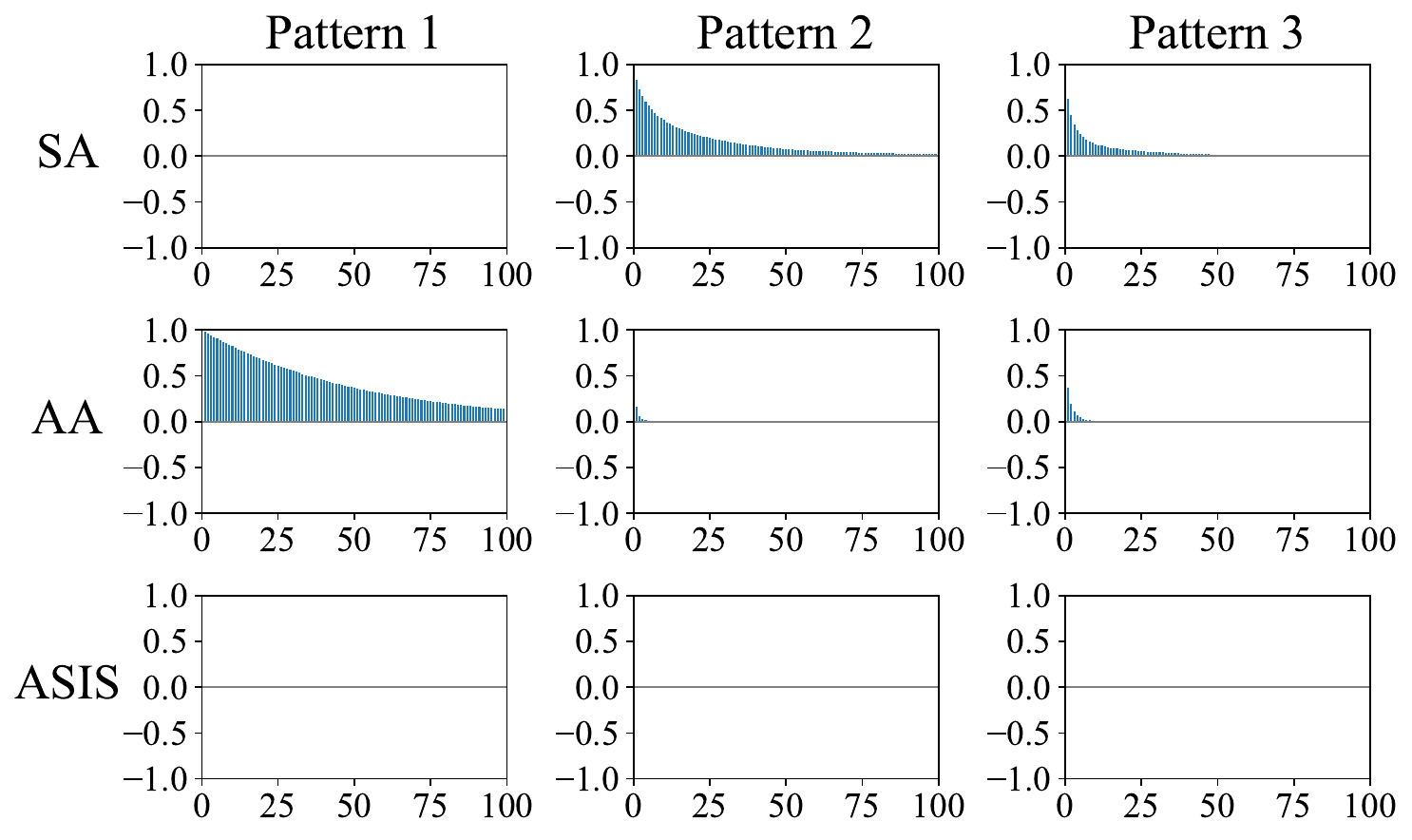}
        \caption{$(N,T) = (10,500)$}
        \label{fig:acf_n10_t500}
    \end{subfigure}
    \hfill
    \begin{subfigure}[t]{0.49\textwidth}
        \centering
        \includegraphics[width=\linewidth]{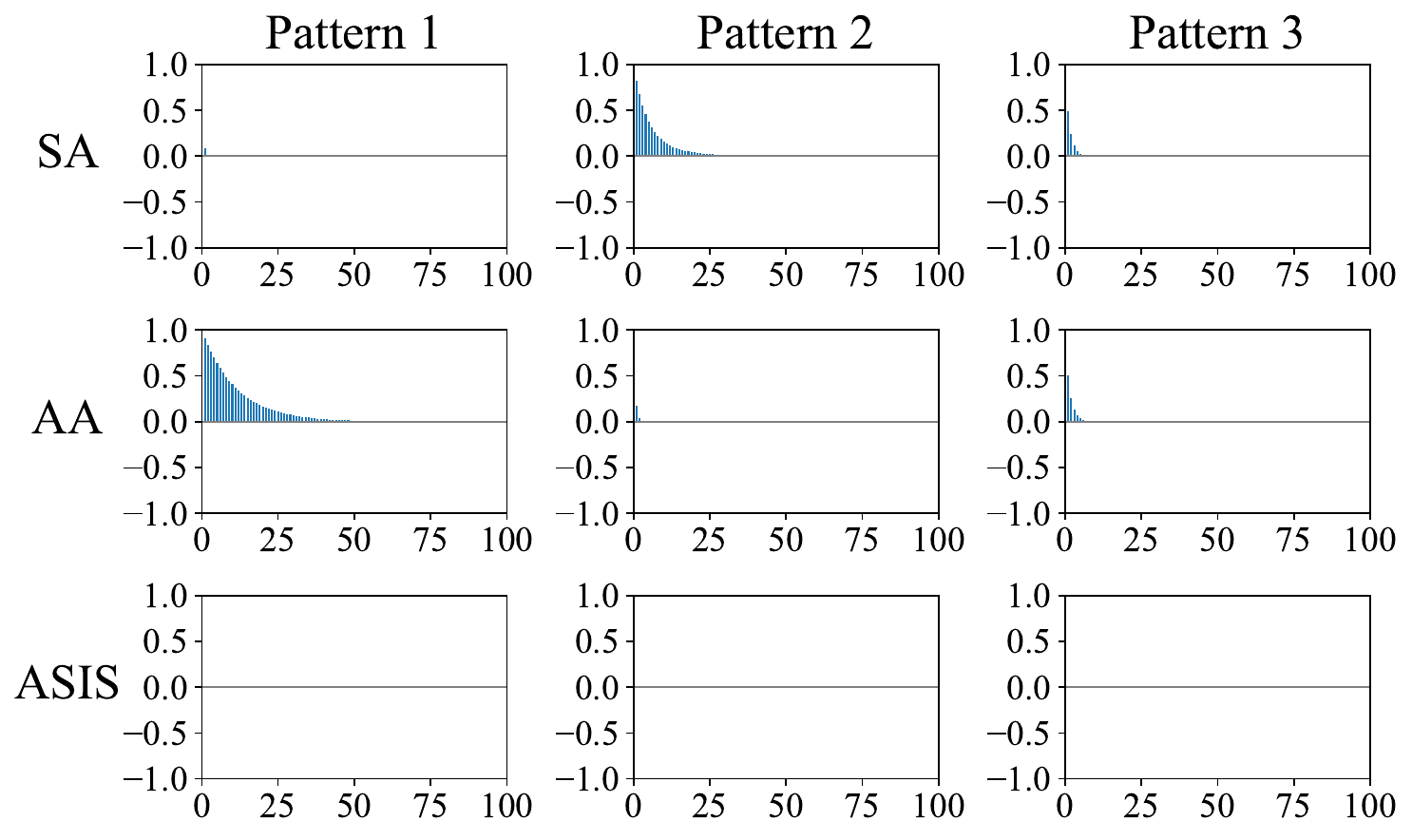}
        \caption{$(N,T) = (500,500)$}
        \label{fig:acf_n500_t500}
    \end{subfigure}

    \caption{ACF plots of $\mu_\alpha$ for the simulation study with six panel configurations}
    \label{fig:2x3_with_captions}
\end{figure}

In \cref{tab:2x3_subtables1,tab:2x3_subtables2,tab:2x3_subtables3}, we compare the Monte Carlo standard errors (MCSEs) of $\mu_\alpha$. 
ASIS consistently showed the best performance across all settings; however, the primary focus of this study is the MCSE ratio between SA and AA under Patterns 1, 2, and 3.
Comparison of Patterns 1 and 2 revealed that SA had consistently lower MCSE than AA in Pattern 1, whereas AA had lower MSCE than SA in Pattern 2. 
These findings matched the predictions of Corollary 1.
Under Pattern 3, SA and AA exhibited noticeable MSCE differences when $N = 10$ (\cref{tab:mcse-comparison1,tab:mcse-comparison2}). However, for $N = 500$ (\cref{tab:mcse-comparison3,tab:mcse-comparison4}), the difference was negligible; these findings also aligned with the theoretical prediction of Corollary 1.
Further support for the theoretical findings from Theorems 1 and 2 is provided in \cref{fig:acf_n10_t10,fig:acf_n10_t100,fig:acf_n10_t500,fig:acf_n500_t10,fig:acf_n500_t100,fig:acf_n500_t500}, where the autocorrelation function (ACF) of $\mu_\alpha$ under ASIS decayed considerably faster than that under SA and AA.

\begin{table}[H]
  \centering

  \begin{subtable}[t]{0.49\textwidth}
  \centering
  \begin{threeparttable}
    \begin{adjustbox}{scale=0.90}
    \begin{tabular}{crrr}
      \toprule
        &
      \multicolumn{1}{c}{SA} &
      \multicolumn{1}{c}{AA} &
      \multicolumn{1}{c}{ASIS} \\
      \midrule
      Pattern 1: \quad  $\sigma_\varepsilon^{2} < T\sigma_\alpha^{2}$ \quad & 2.980 & 6.178 & \bf{2.427} \\ \addlinespace[1pt]
      Pattern 2: \quad $\sigma_\varepsilon^{2} > T\sigma_\alpha^{2}$ \quad & 56.286 & 17.057 & \bf{13.716} \\ \addlinespace[1pt]
      Pattern 3: \quad $\sigma_\varepsilon^{2} = T\sigma_\alpha^{2}$ \quad & 9.644 & 14.399 & \bf{6.697}\\
      \bottomrule
    \end{tabular}
    \end{adjustbox}
    \caption{MCSE comparison of $\mu_\alpha$ ($N = 10, T = 10$)}
  \label{tab:mcse-comparison1}
  \end{threeparttable}
  \end{subtable}
  \hfill
  \begin{subtable}[t]{0.49\textwidth}
  \centering
  \begin{threeparttable}
  \begin{adjustbox}{scale=0.90}
    \begin{tabular}{crrr}
      \toprule
        &
      \multicolumn{1}{c}{SA} &
      \multicolumn{1}{c}{AA} &
      \multicolumn{1}{c}{ASIS} \\
      \midrule
      Pattern 1: \quad  $\sigma_\varepsilon^{2} < T\sigma_\alpha^{2}$ \quad & 0.600 & 2.279 & \bf{0.567} \\ \addlinespace[1pt]
      Pattern 2: \quad  $\sigma_\varepsilon^{2} > T\sigma_\alpha^{2}$ \quad & 10.496 & 1.753 & \bf{1.722} \\ \addlinespace[1pt]
      Pattern 3: \quad  $\sigma_\varepsilon^{2} = T\sigma_\alpha^{2}$ \quad & 1.192 & 1.126 & \bf{0.738} \\
      \bottomrule
    \end{tabular}
    \end{adjustbox}
    \caption{MCSE comparison of $\mu_\alpha$ ($N=500, T=10$) }
  \label{tab:mcse-comparison4}
  \end{threeparttable}
  \end{subtable}
  \caption{MCSE comparison of $\mu_\alpha$ for the simulation data. All values are scaled by $\times 10^{-5}$. Boldface denotes the smallest MCSE. \\ Pattern 1 (SA$<$AA expected)\,: $(\sigma_\varepsilon, \sigma_\alpha) = (1, 1)$, Pattern 2 (SA$>$AA expected)\,: $(\sigma_\varepsilon, \sigma_\alpha) = (10, 1)$, Pattern 3 (SA$=$AA expected)\,: $(\sigma_\varepsilon, \sigma_\alpha) = (\sqrt{10}, 1)$.}
  \label{tab:2x3_subtables1}
\end{table}

\begin{table}[H]
  \begin{subtable}[t]{0.49\textwidth}
  \centering
  \begin{threeparttable}
  \begin{adjustbox}{scale=0.90}
    \begin{tabular}{crrr}
      \toprule
        &
      \multicolumn{1}{c}{SA} &
      \multicolumn{1}{c}{AA} &
      \multicolumn{1}{c}{ASIS} \\
      \midrule
      Pattern 1: \quad  $\sigma_\varepsilon^{2} < T\sigma_\alpha^{2}$ \quad & 3.255 & 8.587 & \bf{2.877} \\ \addlinespace[1pt]
      Pattern 2: \quad  $\sigma_\varepsilon^{2} > T\sigma_\alpha^{2}$ \quad & 53.828 & 17.665 & \bf{13.781} \\ \addlinespace[1pt]
      Pattern 3: \quad  $\sigma_\varepsilon^{2} = T\sigma_\alpha^{2}$ \quad & 9.221 & 13.949 & \bf{6.589} \\
      \bottomrule
    \end{tabular}
    \end{adjustbox}
      \caption{MCSE comparison of $\mu_\alpha$ ($N=10, T=100$)}
  \label{tab:mcse-comparison2}
  \end{threeparttable}
  \end{subtable}
  \hfill
  \begin{subtable}[t]{0.49\textwidth}
  \centering
  \begin{threeparttable}
  \begin{adjustbox}{scale=0.90}
    \begin{tabular}{crrr}
      \toprule
        &
      \multicolumn{1}{c}{SA} &
      \multicolumn{1}{c}{AA} &
      \multicolumn{1}{c}{ASIS} \\
      \midrule
      Pattern 1: \quad  $\sigma_\varepsilon^{2} < T\sigma_\alpha^{2}$ \quad & 0.618 & 2.379 & \bf{0.585} \\ \addlinespace[1pt]
      Pattern 2: \quad  $\sigma_\varepsilon^{2} > T\sigma_\alpha^{2}$ \quad & 8.815 & 1.813 & \bf{1.733} \\ \addlinespace[1pt]
      Pattern 3: \quad  $\sigma_\varepsilon^{2} = T\sigma_\alpha^{2}$ \quad & 1.169 & 1.168 & \bf{0.744} \\
      \bottomrule
    \end{tabular}
    \end{adjustbox}
      \caption{MCSE comparison of $\mu_\alpha$ ($N=500, T=100$)}
  \label{tab:mcse-comparison5}
  \end{threeparttable}
  \end{subtable}
  \caption{MCSE comparison of $\mu_\alpha$ for the simulation data. All values are scaled by $\times 10^{-5}$. Boldface indicates the smallest MCSE. \\
  Pattern 1 (SA$<$AA expected) : $(\sigma_\varepsilon, \sigma_\alpha) = (\sqrt{10}, 1)$, Pattern 2 (SA$>$AA expected) : $(\sigma_\varepsilon, \sigma_\alpha) = (\sqrt{1000}, 1)$, Pattern 3 (SA$=$AA expected) : $(\sigma_\varepsilon, \sigma_\alpha) = (10, 1)$.}
  \label{tab:2x3_subtables2}
\end{table}

\begin{table}[H]
  \begin{subtable}[t]{0.49\textwidth}
  \centering
  \begin{threeparttable}
  \begin{adjustbox}{scale=0.90}
    \begin{tabular}{crrr}
      \toprule
        &
      \multicolumn{1}{c}{SA} &
      \multicolumn{1}{c}{AA} &
      \multicolumn{1}{c}{ASIS} \\
      \midrule
      Pattern 1: \quad  $\sigma_\varepsilon^{2} < T\sigma_\alpha^{2}$ \quad & 8.645 & 57.364 & \bf{8.598} \\ \addlinespace[1pt]
      Pattern 2: \quad  $\sigma_\varepsilon^{2} > T\sigma_\alpha^{2}$ \quad & 39.168 & 10.528 & \bf{9.015} \\ \addlinespace[1pt]
      Pattern 3: \quad  $\sigma_\varepsilon^{2} = T\sigma_\alpha^{2}$ \quad & 12.861 & 7.040 & \bf{4.687} \\
      \bottomrule
    \end{tabular}
    \end{adjustbox}
      \caption{MCSE comparison of $\mu_\alpha$ ($N=10, T=500$)}
  \label{tab:mcse-comparison3}
  \end{threeparttable}
  \end{subtable}
  \hfill
  \begin{subtable}[t]{0.49\textwidth}
  \centering
  \begin{threeparttable}
  \begin{adjustbox}{scale=0.90}
    \begin{tabular}{crrr}
      \toprule
        &
      \multicolumn{1}{c}{SA} &
      \multicolumn{1}{c}{AA} &
      \multicolumn{1}{c}{ASIS} \\
      \midrule
      Pattern 1: \quad  $\sigma_\varepsilon^{2} < T\sigma_\alpha^{2}$ \quad & 0.607 & 2.289 & \bf{0.573} \\ \addlinespace[1pt]
      Pattern 2: \quad  $\sigma_\varepsilon^{2} > T\sigma_\alpha^{2}$ \quad & 3.746 & 1.493 & \bf{1.303} \\ \addlinespace[1pt]
      Pattern 3: \quad  $\sigma_\varepsilon^{2} = T\sigma_\alpha^{2}$ \quad & 1.165 & 1.199 & \bf{0.753} \\
      \bottomrule
    \end{tabular}
    \end{adjustbox}
    \caption{MCSE comparison of $\mu_\alpha$ ($N=500, T=500$)}
    \label{tab:mcse-comparison6}
  \end{threeparttable}
  \end{subtable}
  \caption{MCSE comparison of $\mu_\alpha$ for the simulation data. All values are scaled by $\times 10^{-5}$. Boldface indicates the smallest MCSE.\\ Pattern 1 (SA$<$AA expected) : $(\sigma_\varepsilon, \sigma_\alpha) = (\sqrt{50}, 1)$, Pattern 2 (SA$>$AA expected) : $(\sigma_\varepsilon, \sigma_\alpha) = (50, 1)$, Pattern 3 (SA$=$AA expected) : $(\sigma_\varepsilon, \sigma_\alpha) = (\sqrt{500}, 1)$.}
  \label{tab:2x3_subtables3}
\end{table}

\section{Real Data Analysis}
The proposed algorithm was demonstrated on the U.S. ``Cigarette'' panel dataset, originally compiled by \cite{baltagi1986estimating} and later popularized by \cite{stock2003introduction}. 
This dataset covers 48 states over 11 years and is widely used as a standard benchmark for teaching and empirical evaluation in econometrics. 
It is widely featured in leading textbooks \cite{kleiber2008applied} and instructional materials such as \cite{fritsch2024teaching}, and is readily available in R and Stata, ensuring full replicability. 
Since benchmark estimates---such as elasticities and other key parameters---are firmly established in the literature, the numerical and computational advantages of the proposed algorithm can be meaningfully assessed against these well-known reference values; this enabled a focused evaluation of the performance of the proposed algorithm.

Following \cite{baltagi1986estimating,stock2003introduction}, a panel data regression model was constructed for per capita cigarette pack consumption.
As real-world panel datasets almost always include influences beyond individual-specific effects, controlling for observable covariates is essential.
Therefore, real per capita income, average retail price per pack (including sales taxes), and average excise tax per pack were included as covariates herein.

Thus the panel data regression model to be estimated is
\begin{align*}
  \log(\text{demand}_{it})
  =\alpha_i+\beta_1 \log(\text{income}_{it})
  +\beta_2 \log(\text{price}_{it}) +\beta_3 \log(\text{tax}_{it}) +\varepsilon_{it},
  \quad\varepsilon_{it}\sim\mathcal{N}\!\bigl(0,\sigma_\varepsilon^{2}\bigr),
\end{align*}
where \(i\) indexes states and \(t\) indexes years.  

Define
\begin{align*}
    y_{it} &= \log(\text{demand}_{it}), \\
    \mathbf{x}_{it} &= [\log(\text{income}_{it})\ \log(\text{price}_{it})\ \log(\text{tax}_{it})]^{\intercal}, \\
    \boldsymbol{\beta} &= [\beta_1\ \beta_2\ \beta_3]^{\intercal}.
\end{align*}
By inserting $\mathbf{x}_{it}^{\intercal}\boldsymbol{\beta}$ on the right-hand side of \cref{sa,aa}, we obtain
\begin{align*}
  \text{(SA)}\quad
  y_{it} &= \alpha_{i} + \mathbf{x}_{it}^{\intercal}\boldsymbol{\beta} + \epsilon_{it},\quad \alpha_{i} \sim \mathcal{N}(\mu_{\alpha},\sigma_{\alpha}^{2}), \\
  \label{aa2}
  \text{(AA)}\quad
  y_{it} &= \mu_{\alpha} + \tilde{\alpha}_{i} + \mathbf{x}_{it}^{\intercal}\boldsymbol{\beta} +\epsilon_{it},\quad \tilde{\alpha}_{i} \sim \mathcal{N}(0,\sigma_{\alpha}^{2}), \\
  \epsilon_{it} &\sim \mathcal{N}(0,\sigma_{\epsilon}^{2}),\quad
  i =1,\dots, N,\quad t = 1,\dots, T. \nonumber
\end{align*}

Although introducing $\boldsymbol{\beta}$ increases model complexity, the term $\mathbf{x}_{it}^{\intercal} \boldsymbol{\beta}$ allows part of $y_{it}$ to be explained by observable factors.
As a result, the level of the unobservable individual effect $\alpha_i$ may shift, and thus the levels of $\mu_\alpha$ and 
$\sigma^2_\alpha$ may also change. 
However, as explained in the introduction, by partialing them out as $\tilde{y}_{it}=y_{it}-\mathbf{x}_{it}^{\intercal}\boldsymbol{\beta}$, we recover an individual‐effect‐only structure for \(\alpha_i\); hence the efficiency results proved for that setting still hold.
Through this real data analysis, we also demonstrate that Theorems \ref{thm:theorem1} and \ref{thm:theorem2}, as well as Corollary \ref{cor:corollary1}, still hold even when $\mathbf{x}_{it}^{\intercal}\boldsymbol{\beta}$ is included in the model.

Prior distributions for $(\alpha_1,\dots,\alpha_{N},\boldsymbol{\beta},\mu_\alpha,\sigma_\alpha,\sigma_\varepsilon)$ are assigned as
\begin{align*}
\alpha_i &\sim \mathcal{N}\bigl(\mu_\alpha,\sigma_{\alpha}^{2}\bigr),\quad
  \mu_{\alpha}\sim \mathcal{N}\bigl(0,100\bigr),\quad
  \sigma_{\alpha}\sim \mathcal{C}^+\bigl(0,10\bigr), \quad
\boldsymbol{\beta}&\sim\mathcal{N}\bigl(\boldsymbol{0},100\boldsymbol{I}\bigr),\quad
  \sigma_\varepsilon\sim\mathcal{C}^+\bigl(0,10),
\end{align*}
where $\mathcal{C}^{+}$ denotes a half-Cauchy distribution and $\boldsymbol{I}$ denotes the identity matrix.
All parameters can be readily generated from their joint posterior distribution via a Gibbs sampler coupled with ASIS.
See \cite{nakakita2023hierarchical} among others for more details on the sampling algorithm.

\begin{figure}[H]
  \centering
  \includegraphics[width=1.0\linewidth]{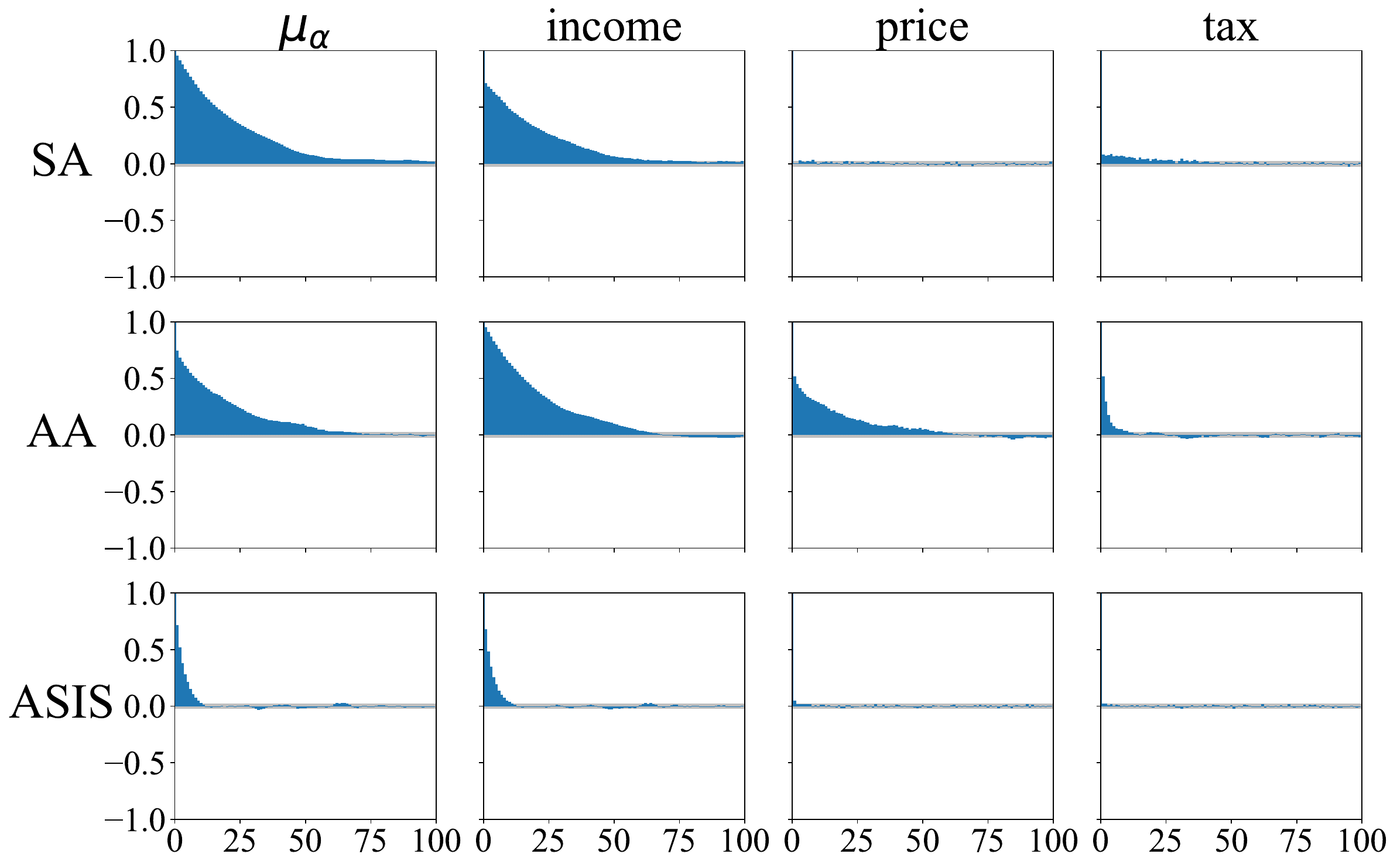}
  \caption{ACF plots for the cigarette data.}
  \label{fig:fig_cigarette_acf}
\end{figure}

\cref{fig:fig_cigarette_acf,tab:mcse-comparison-cigarette} report the empirical results. 
First, we investigate the global mean $\mu_\alpha$. 
Across ACF and MCSE metrics, ASIS consistently outperforms SA and AA. 
The lower ACF and MCSE values for ASIS confirm its superior mixing and sampling efficiency on real data, consistent with the improvements observed in the simulation study.

Next, we examine the sampling efficiency of \( \boldsymbol{\beta} \). 
For \( \beta_1 \), ACFs under both SA and AA parameterizations exhibit slow decay, remaining substantial even at longer lags. 
In contrast, ASIS demonstrates a markedly faster decay of the ACF, even at short lags, indicating more efficient mixing.
As for \( \beta_2 \), while both ASIS and SA yield ACFs that rapidly approach zero, AA fails to achieve similar convergence, with autocorrelations persisting across lags.
This empirical finding contradicts the conventional wisdom established by \cite{gelfandEfficientParametrisationsNormal1995}, which advocates the general superiority of AA. 
In contrast, our study provides a theoretical foundation for the consistent efficiency of ASIS and further identifies the mathematical conditions under which SA outperforms AA. 
These theoretical insights are corroborated by the real data analysis.
Lastly, for \( \beta_3 \), we observe a similar pattern: ASIS yields the lowest ACFs, achieving the most efficient posterior sampling, while SA again outperforms AA. 
These results reinforce the robustness of our theoretical claims across multiple regression coefficients.

The conclusions drawn from the ACFs are further corroborated by the MCSEs.  
As shown in \cref{tab:mcse-comparison-cigarette}, ASIS achieves the lowest MCSEs not only for \( \mu_\alpha \), but also for \( \beta_1 \) and \( \beta_3 \), indicating superior sampling efficiency.  
Although SA achieves the lowest MCSE for \( \beta_2 \), the difference relative to ASIS is numerically small and statistically insignificant.
Moreover, ASIS still substantially outperforms AA in this case, preserving its overall advantage.  
Overall, these findings indicate that ASIS improves sampling efficiency across both \( \mu_\alpha \) and the regression coefficients \( \boldsymbol{\beta} \), thereby validating its utility as an effective reparameterization method.

\begin{table}[H]
  \centering
  \begin{threeparttable}
    \begin{tabular*}{1\linewidth}{@{\extracolsep{\fill}}ll cccc}
      \toprule
      & Method & SA & AA & ASIS & \\
      \midrule
      & $\mu_\alpha$ & 8.673 & 8.232 & \bf{3.072} & \\
      & $\beta_1$ (income) & 3.674 & 4.183 & \bf{1.471} & \\
      & $\beta_2$ (price) & \bf{0.206} & 0.813 & 0.211 & \\
      & $\beta_3$ (tax) & 0.710 & 0.657 & \bf{0.349} & \\      
      \bottomrule
    \end{tabular*}
  \caption{MCSE comparison of $\mu_\alpha$ and $\boldsymbol{\beta}$ for the cigarette data. All values are scaled by $\times 10^{-3}$. Boldface denotes the smallest MCSE.}
  \label{tab:mcse-comparison-cigarette}
  \end{threeparttable}
\end{table}

\section{Conclusion}
\label{sec:conclusion}
This study has advanced Bayesian panel data analysis in three key respects.
First, it provides the first rigorous theoretical justification for applying ASIS to hierarchical panel models. 
The asymptotic theory demonstrates that when the product of the prior variance of the unobserved heterogeneity and $N$ is sufficiently large, the latent individual effects~$\alpha_i$ can be sampled almost independently of the global mean $\mu_{\alpha}$. 
This decoupling explains why ASIS, already effective in state-space models, continues to mix efficiently in panel settings, highlighting its suitability for modern ``tall'' datasets.

Second, we introduce clear and easy-to-verify criteria for determining whether SA or AA yields faster geometric convergence. 
We also show that ASIS, which interweaves SA and AA, renders the sequence $\{\mu_{\alpha}^{(r)}\}_{r=1}^{R}$ approximately IID and achieves optimal convergence efficiency in all cases.

Third, although the theoretical results are asymptotic, the simulation study confirms that the ordering predicted by Corollary 1 holds even for small panels \(e.g., N=10\). This robustness extends the practical applicability of the method, which has been successfully applied in empirical studies across finance \cite{Nakakita_2021}, marketing \cite{Saito_2024}, and sports analytics \cite{nakakita2023hierarchical}.

Regardless, several aspects require further exploration beyond the scope of this study.  
As the present analysis focused on the simplest baseline specification, incorporating additional regressors or increasingly complex hierarchical structures may influence certain quantitative aspects; however, the qualitative ordering between SA and AA will likely remain robust.
Extending the theoretical framework and sampler to non-Gaussian panel models, such as panel logit or probit, will substantially enhance their empirical applicability.
Finally, many real-world datasets exhibit cross-sectional heterogeneity and rich temporal dynamics. 
Adapting the proposed approach to time-series hierarchies is a promising future research direction.

\bibliographystyle{apa}
\bibliography{panelASIS}
\end{document}